\documentclass[journal]{IEEEtran}
\ifCLASSINFOpdf
   \usepackage[pdftex]{graphicx}
\else
\fi

\usepackage{cite}

\ifCLASSOPTIONcompsoc
  \usepackage[caption=false,font=normalsize,labelfont=sf,textfont=sf]{subfig}
\else
  \usepackage[caption=false,font=footnotesize]{subfig}
\fi

\usepackage{stfloats}
\begin{document}

\title{A Trusted and Privacy-preserving Internet of Mobile Energy}
\author{
% \\ \\ \\ \\ \\
\IEEEauthorblockN{
    Raja Jurdak\IEEEauthorrefmark{1},
    Ali Dorri\IEEEauthorrefmark{1},
    Mahinda Vilathgamuwa\IEEEauthorrefmark{3}\\
}
\IEEEauthorblockA{
    \IEEEauthorrefmark{1}Trusted Networks Lab, School of Computer Science, QUT\\
    \IEEEauthorrefmark{3}School of Electrical Engineering and Robotics, QUT\\
    \{r.jurdak,ali.dorri,mahinda.vilathgamuwa\}@qut.edu.au}
    \thanks{\copyright 2021 IEEE. Personal use of this material is permitted. Permission from IEEE must be obtained for all other uses, in any current or future media, including reprinting/republishing this material for advertising or promotional purposes, creating new collective works, for resale or redistribution to servers or lists, or reuse of any copyrighted component of this work in other works.}
}

%\markboth{}

\maketitle

\begin{abstract}
The rapid growth in distributed energy sources on power grids leads to increasingly decentralised energy management systems for the prediction of power supply and demand and the dynamic setting of an energy price signal. Within this emerging smart grid paradigm, electric vehicles can serve as consumers, transporters, and providers of energy  through two-way charging stations, which highlights a critical feedback loop between the movement patterns of these vehicles and the state of the energy grid. This paper proposes a vision for an Internet of Mobile Energy (IoME), where energy and information flow seamlessly across the power and transport sectors to enhance the grid stability and end user welfare. We identify the key challenges of trust, scalability, and privacy, particularly location and energy linking privacy for EV owners, for realising the IoME vision. We propose an information architecture for IoME that uses scalable blockchain to provide energy data integrity and authenticity, and introduces one-time keys for public EV transactions  and a verifiable anonymous trip extraction method for EV users to share their trip data while protecting their location privacy. We present an example scenario that details the seamless and closed loop information flow across the energy and transport sectors, along with a blockchain design and transaction vocabulary for trusted decentralised transactions. We finally discuss the open challenges presented by IoME that can unlock significant benefits to grid stability, innovation, and end user welfare. 
\end{abstract}


\IEEEpeerreviewmaketitle

\section{Introduction}
The emergence of the two-way communication model and Distributed Energy Sources (DES) is transforming traditional power systems from largely centralised energy production  to more decentralised and connected management systems, where all nodes can inject energy and communicate with other parties involved. This is called the smart grid. A broad range of devices connected through the smart grid can inject information to the network and communicate with the energy manager which in turn leads to the emergence of new services to facilitate energy management and prevent energy shortage and price fluctuation.

Electric vehicles (EVs) are emerging as new unconventional and highly disruptive participants in the grid that can add significant benefit and flexibility.  EVs are conventionally viewed as energy consumers that are charged at charging stations to support their movement. Notably, EVs are equipped with a high capacity battery, typically an order of magnitude larger than home batteries,  that stores energy to power the vehicle. EV batteries, coupled with the recent introduction of two-way charging/discharging stations~\cite{Morrissey2016}, enable EVs to also serve as mobile energy transporters within an electrical grid and as energy suppliers when they have disposable energy. Studies are conducted to explore the effectiveness of EV batteries as mobile storage~\cite{ARENA} and as energy transporters~\cite{Kosmanos18}  using vehicle-to-vehicle wireless charging and EV route optimization based on the availability of vehicles and static charging stations. Such functionalities allow EVs to contribute to power quality improvement within specific zones of the grid. 

%Consider a scenario where a specific region of the grid is experiencing a peak in energy demand, where transmission lines are at maximum capacity for energy transfer. While demand still outstrips supply,  further increasing the capacity of transmission lines is not an option as it will cause grid instability leading to complete shutdown in the area. EV's within this region that have sufficient charge in their batteries can opt to inject energy into this region's grid or directly to consumers, through a two-way charging station, in exchange for monetary gain. This can help the grid better cope with the challenges of peak demand.  

The trends toward greater forecasting in smart grids, distributed energy generation and greater adoption and charging/discharging flexibility of EVs highlight a greater convergence between the energy and transport sectors. This vision can lead us to a future with closed loop information and energy flow between the grid and electric vehicles, which will benefit both the grid stability and the end user interests. We refer to this vision as the Internet of Mobile Energy (IoME).

 Achieving this vision involves three key challenges: (1) trust; (2) privacy; and (3) scalability. On trust, it is critical to ensure the integrity and authenticity of energy transactions, involving both fixed smart grid participants as well as EVs, as these transactions will determine the current and future energy state of the system. A related trust challenge is to ensure the veracity of contributed mobility patterns by EVs to inform smart grid energy predictions.  Regarding privacy, smart grids conventionally reveal a participant’s energy usage and generation which can help an attacker infer the participant’s activity. In IoME, the user’s mobility patterns, in addition to their energy supply and usage, can be inferred by linking their EV’s charging/discharging locations to their identity, leading to both location and activity privacy risks. Finally, scalability is critical as IoME networks will range from city scale to country scale, potentially involving millions of participating energy nodes and their associated transactions. Existing architectures~\cite{Lopez18} support energy transactions and information exchange within an individual sector, so they do not address the above challenges of cross-sectoral energy and information sharing.

%The mobility of EVs and the use of their mobility patterns to inform energy forecasts introduces unique challenges around privacy, trust, and security. On privacy, an EV that transacts energy at a charging station already reveals its license plate and its current location. For the vehicle owner to maintain anonymity in the energy network, she must rely on changeable pseudonyms that prevent an attacker from linking her digital identity to the vehicle. This can allow the vehicle partner to anonymously pay for/receive funds for transacted energy. On trust, payments for energy transactions by EVs and charging stations must honour the actual energy transferred, and this is challenging particularly in the presence of changeable pseudonyms by the EVs that may try to evade payments. Therefore, how to ensure full trust and accountability in the network without revealing the identities of the EV owners is another major challenge. Finally, Scalability is another major challenge, with the need to support city-scale transactions from mobile EVs and charging stations. Conventional energy management frameworks rely on centralized brokered communication model which in turn creates a single point of failure and many-to-one traffic challenges.

To address the scalability challenge requires a decentralised information architecture that can eliminate the single point of failure. This architecture must support  security, privacy, and trust while providing sufficient transparency for effective energy transactions. In recent years, there has been extensive research toward addressing the above challenges using blockchain technology~\cite{Miglani20}. Blockchain has attracted tremendous attention due to its salient features which includes decentralization, security, anonymity, auditability and transparency ~\cite{Dorri_SPB,Farao19}.  Blockchain is a shared ledger of blocks where each block stores records of the communications between the participating nodes, also known as transactions. Each node is known by a  public key (PK) that is used as pseudonym. Recent lightweight blockchains~\cite{Dorri19,Divya18}  promote scalability, increase transaction throughput, and reduce delay over conventional blockchains.

%The limited data availability challenge (challenge (3)) is being addressed through the increased reliance on IoT devices to provide deeper context for determining energy predictions.  To address the data processing challenge (challenge (4)) ML algorithms~\cite{Luo19}, enable the energy managers to process data collected from heterogeneous sources and predict the future energy demand based on the history and current requirements of the nodes. Many of these prediction algorithms rely not only on energy grid data sources, but also on IoT sensor data to predict microclimate patterns~\cite{Saber17}, as well as external weather data to predict macroclimate~\cite{Li17}, which both correlate with energy availability from renewable energy sources such as roof-top solar. While  technology innovation is already disrupting smart grids, much of it is occurring in silos of activity which only considers the impact of either blockchain or ML in energy management. The convergence and tighter coupling of these technologies can provide a key step towards achieving the IoME vision. 

\begin{figure}
    \centering
    \includegraphics[width=10cm ,height=6cm,keepaspectratio]{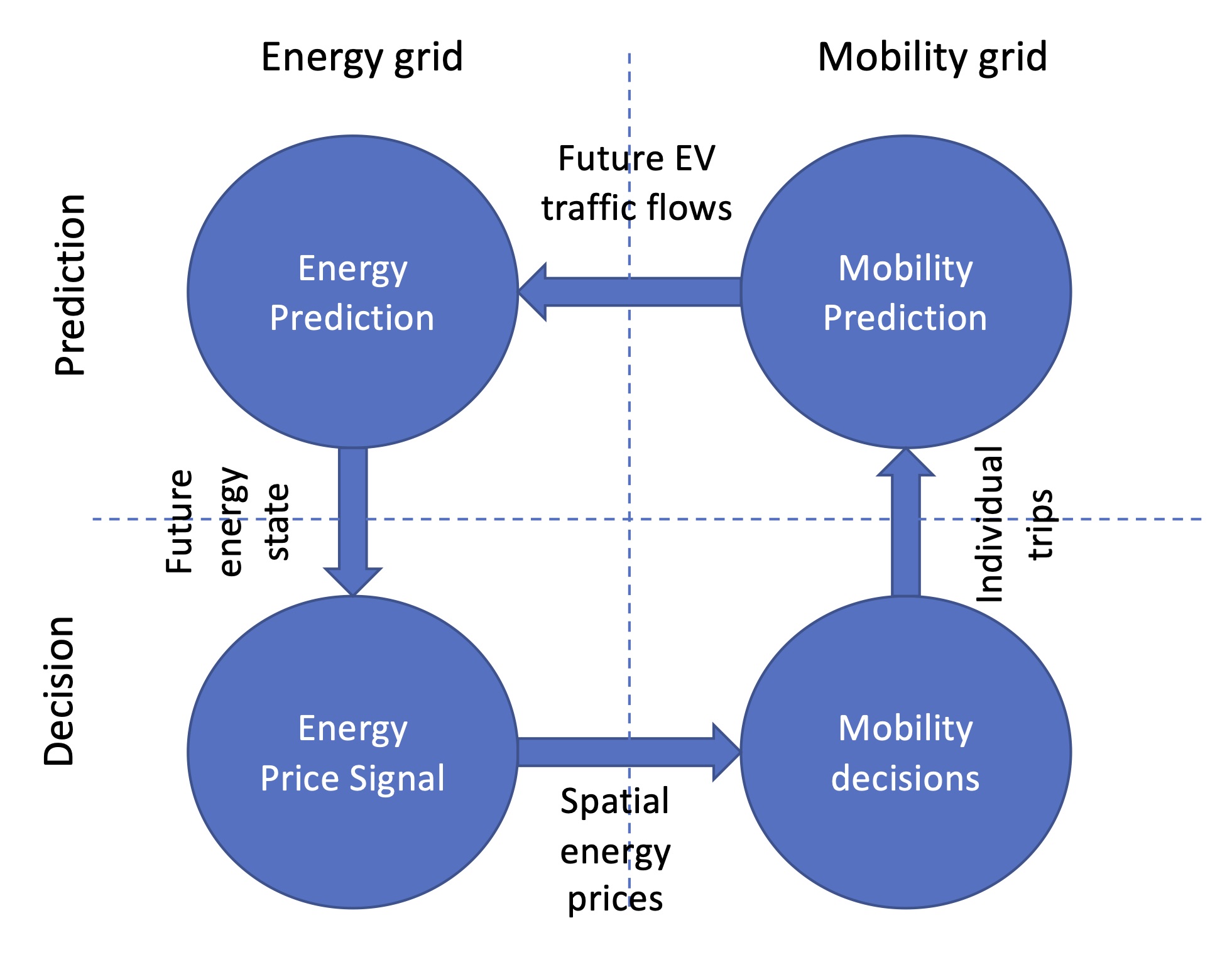}
    \caption{Information flow in IoME}
    \label{fig:IoME_info_flow}
\end{figure}
In this paper, we propose a novel IoME architecture underpinned by scalable blockchains to facilitate energy management. The IoME architecture creates a feedback loop between the energy and transport grids by considering the mobility patterns of EVs in predicting and identifying energy prices, as shown in Fig.~\ref{fig:IoME_info_flow}. The IoME architecture enables all participants to communicate and share information in a trusted, secure and anonymous manner. The participating nodes record their energy supply or demand information on blockchain, ensuring data integrity, authenticity, and non-repudiation which enables trusted prediction of the future demand and supply of energy  in the network. Energy prediction in the IoME architecture also utilizes the mobility patterns of the EVs as an input which enhances the accuracy of the energy demand or supply by considering the mobile nodes in the network.   The data is processed using machine learning (ML) algorithms and the result is announced to the network as an energy price signal.  The security of IoME is inherited from blockchain's security mechanisms.

To address the privacy challenge in IoME, particularly attacks that attempt to link to EV owners' historical transactions, the IoME architecture enables EV's to generate one-time public keys. Whenever the EV docks at a charging station, it shares a one-time key with the charging station. The station signs the key, along with the current timestamp.  This prevents Sybil attacks as the station can confirm the presence of the vehicle at its location when the key is received before signing the key/timestamp pair. These one-time keys are separate from all keys used by the EV for any private transactions, such as in the home. The use of one-time key pairs for each trip in public charging transactions ensures that EVs cannot be individually tracked and that their charging transactions cannot be linked over time, while enabling them to reliably transact within the IoME network. 

Recall that the mobility patterns must be fed back into energy predictions as the EV movements will impact the future energy state. High resolution mobility patterns require construction of an origin-destination (O-D) matrix to capture traffic flows. The O-D matrix relies on aggregation of individual trip information from EVs, yet the use of our one-time keys prevents direct extraction of individual EV trips as the keys used at a trip origin are different from the destination keys for this EV. To ensure the mobility patterns are trusted, we introduce a verifiable and anonymous trip extraction method using signed verification of presence at each charging station to verify that the EV was indeed at the origin and destination charging stations at the claimed times, without revealing any EV identifiable information. We next discuss the IoME information flows, before describing the blockchain design through an example scenario.

\section{Internet of Mobile Energy} \label{sec:emergingEnergyMobility}

\begin{figure*}
    \centering
    \includegraphics[width=13cm ,height=10cm,keepaspectratio]{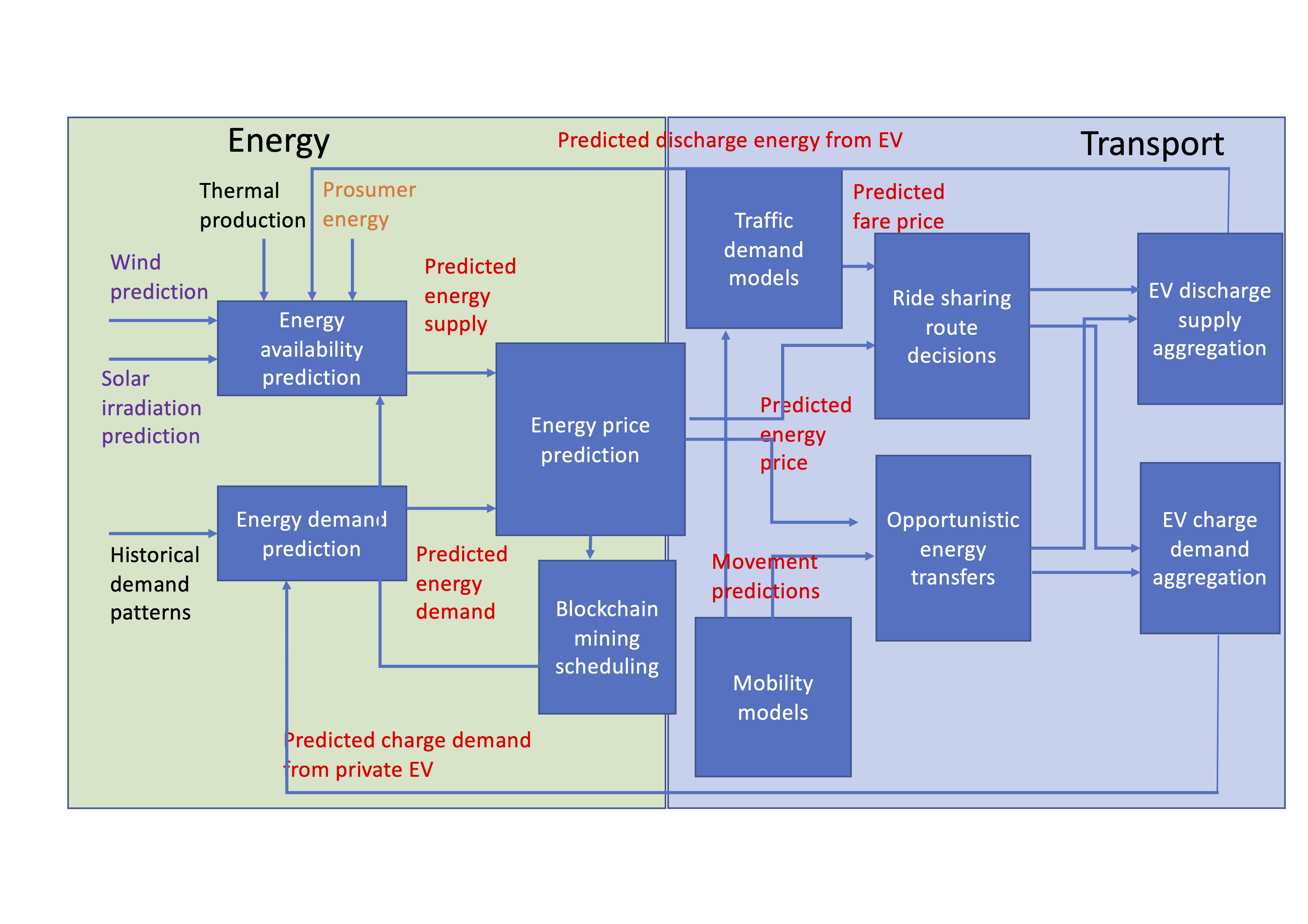}
    \caption{Combining energy prediction with the mobility of the smart vehicles.}
    \label{fig:mobility}
\end{figure*}
\subsection{IoME vision}
\label{Sec:EnergyMobility}
The IoME future relies on seamless interaction between energy and transport agents.  Fig. \ref{fig:IoME_info_flow} shows a high-level overview of IoME, in which the data of the IoT devices, smart grid participants, and EVs flow freely forming a feedback loop, as shown in Fig.~\ref{fig:mobility}. Energy demand and supply predictions, based on grid, IoT, and weather data, determine the energy price signal. The price signal in turn will drive the mobility decisions of EVs within the transport network. In the mobility grid, mobility patterns are seeded by mobility models, traffic demand models, and any historical data. These models, along with the energy price signal, determine the mobility decision of ride sharing fleets as well as private vehicle owners to opportunistically transfer energy for profit. The mobility decisions in turn determine prospective spatiotemporal charging and discharging patterns of EVs, which then feedback into energy supply and demand prediction. 

We now discuss these steps in more detail, starting with the energy grid, followed by the mobility grid.  

\textit{\textbf{Energy grid}}

The main aims here are: (1) to predict the energy demand and supply in the network; (2) to provide pricing signals to balance  the load in the grid, maintaining the grid's operational stability. We refer to the nodes that run the learning algorithm as learners. The energy grid operations within IoME involve energy prediction and  energy price signal generation.

\textbf{Energy Supply and Demand Prediction}:  To predict the energy demand and supply, the learners must gather data from IoT devices, transport nodes and energy nodes. The IoT devices capture various contextual data (DESs, weather) and daily schedule of users which can feed into the machine learning algorithm to enhance the prediction accuracy. Prediction of energy supply and demand can occur at building energy managers for reduced grid overhead.
    
    A key contribution of this paper is to consider the EVs' mobility patterns in predicting the energy supply or demand. EVs can charge  in a charging station when they need energy or discharge when they have disposable energy, highlighting the need to incorporate EV mobility patterns in energy state predictions. EV Mobility data  can be sourced through individuals or organisations opting in for receiving real-time energy pricing signals that can allow them to guide their movement decisions.  The IoME architecture must provide trustworthy mobility patterns while preserving the privacy of individual EV owners, which we address through  one-time keys and verifiable anonymous trip extraction.
    
    A critical data source that informs mobility \emph{and} energy potential is the state of health (SOH) of EV batteries. The SOH determines the amount of energy each battery can contribute, and the range of reachable charging stations based on its energy data. This group of charging stations constrains the search space for possible routes and destinations for each EV. All such data can be fused for the purposes of mobility modeling and prediction. The grid manager (GM) uses the mobility predictions, alongside the energy supply and demand predictions, to forecast the overall energy state in the smart grid.

   \textbf{Energy Price Signal}:      Based on the predicted energy demand and supply, the GM identifies the energy price and releases price signals to the network. When demand is large relative to supply, the energy price increases and vice versa. The data processing may happen in a hierarchical fashion. This feature enables energy prices to be region-specific depending on the balance of their local energy supply and demand, which will impact the mobility pattern as discussed below.     The participating nodes may alter their energy usage patterns according to the price signal. \\
%\end{itemize}

\textit{\textbf{Mobility Grid}}

The location-specific energy price signal that depends on supply/demand conditions, or grid power quality considerations,  is  used by participants in the transport sector, mainly EVs. Energy managers of the distribution system may offer incentives to EVs to transfer energy from one location to another. Suppose a suburb whose distribution lines  serve as an interconnection for other surrounding suburbs has its distribution lines overloaded. EV's can be incentivised to visit the surrounding suburbs and to discharge energy in order to relieve the load on the loaded lines. 

In conventional energy trading platforms, the energy producer  must inject  energy to the grid in exchange for a feed-in tariff, which is typically lower than the energy price that consumers pay for buying energy from the grid. Therefore, such trading platforms reduce the benefits of the energy producers, and may create power quality deterioration situations as outlined above. To address this challenge, EVs can be guided to transfer energy to the consumer directly to increase their economic gains. %For instance, an EV may discharge its battery in a two-way charging station in busy suburbs where there is high energy demand in order to increase its gain. 

The  benefits of this approach for the transport sector  are two-fold. EV's can consider the energy price signals and their mobility context to either charge/discharge with low/high price, or function as energy transporters to transfer energy from one location to another location. This can help to improve the power quality in the transmission/distribution system and reduces the chance of power shutdown. Second, ride sharing companies can feed pricing signals to their route decision making  and  price estimation algorithms to make their fleet of vehicles aligned with the market energy demand or supply and thus increase their revenue. The drivers may also consider the benefit gained from selling/buying/transporting energy  while deciding to choose a ride. 

Individual mobility \textit{decisions} are aggregated to drive mobility \textit{predictions}, which then feed back into energy supply and demand prediction to close the loop in IoME. Reliance on individual trips to create an O-D matrix would typically risk the location privacy of users. In IoME, we allow vehicles to use a pair of one-time keys at each charging station, which protects their location privacy, but makes trip extraction challenging as the keys at source and destination stations cannot be matched. This motivates our verifiable yet anonymous trip extraction.  

Realising the IoME vision requires an information infrastructure to support accurate prediction of future energy and mobility states and enable seamless information flow across all participants within the energy and transport grids. We propose the IoME information architecture next. 

%This section outlines the proposed energy mobility framework. Figure \ref{fig:my_label} shows a high-level overview of the proposed framework. In the proposed framework, the data of the IoT devices and smart grid participants are fed into a learning algorithm that predicts the future energy supply and demand to determine the price signal as shown in Figure \ref{fig:mobility}. Our platform connects energy  and the transportation sectors. Electric Vehicles (EVs) and SPs in the transportation industry, e.g., ride sharing companies, receive the pricing signal and decide on proper time to charge/discharge, optimized routes toward a destination to maximize the benefit of the user, and receive benefits for transferring energy.    We first outline an overview of the proposed framework in Section \ref{sec:overview} .... 

\subsection{The IoME information architecture} 
\label{sec:infoarch}
\begin{figure*}
    \centering
    \includegraphics[width=12cm ,height=8cm,keepaspectratio]{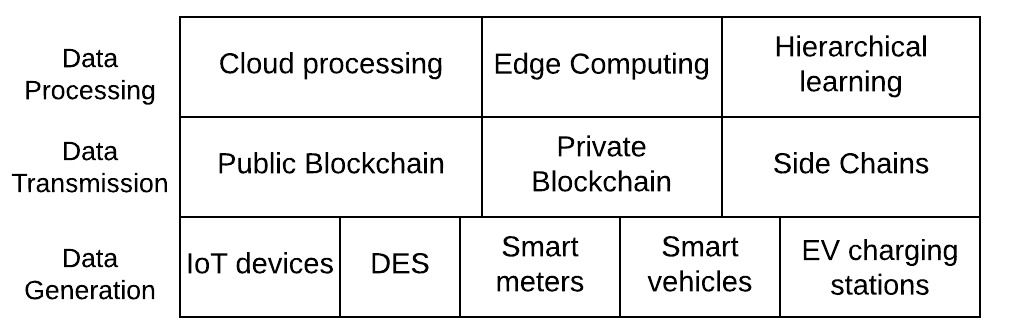}
    \caption{Three layers of the proposed architecture.}
    \label{fig:layers}
\end{figure*}

The IoME information architecture encompasses  heterogeneous devices from energy and transport sectors that perceive the environment and transfer the data to the SPs to offer personalized automated services to the users. The proposed framework consists of three  layers which are summarized in Fig.~\ref{fig:layers} and are described below.

\emph{The Data Generation Layer}: This layer consists of a broad range of heterogeneous devices that sense the environment and send the resulting data to the processing layer through the communication layer. The data generation sources include: i) smart grid nodes: renewable energy sources, e.g. solar panels, traditional energy producers,  transmission infrastructure, EVs, and energy companies, ii) transport nodes: including road side infrastructure,  city managers, vehicle manufacturers, and vehicle service centers, and iii) IoT devices: including microclimate sensors,  smart phones, and smart appliances.

\emph{The Data Communication Layer}:  Blockchain is employed as the underlying trusted communication layer where the participants can exchange data or control messages. Each participating node is known by a changeable Public Key (PK) that introduces a level of anonymity. The IoME architecture uses a single public blockchain as the main chain that connects participants either directly or indirectly, i.e. through a gateway or controller. The purpose of the public chain is to ensure transparency and auditability in the IoME trading ecosystem, among entities that may not trust each other. Within the IoME ecosystem, groups of participants may indeed trust each other, such as vehicle fleets in a ride sharing service, virtual power plant  or microgrid participants. The IoME architecture supports private or side chains, depending on the application, to share data within these groups which in turn increases their privacy and reduces delay and overhead in managing the main chain. We refer to such chains as child chains. All the child chains are connected to the public chain where their hashes  are stored periodically.  Note that any transaction from a child chain participant to a non-participant must go through the main chain via the relevant child chain manager, due to the lack of trust outside the scope of the private chains. 

An EV in IoME employs two one-time PKs that are used exclusively for public charging or discharging during a single trip. These dedicated one-time keys maintain unlinkability with the vehicle owner's other energy transactions, such as within a home or an office, protecting the owner's location privacy.

\emph{The Data Processing Layer}: In this layer, the data produced in layer 1 is processed in the cloud, at edge devices, or hierarchically across the network. A key  privacy consideration at this layer is how to anonymously extract verifiable trips from EV for inclusion in an aggregated O-D matrix, to complete the IoME feedback loop. The EV owner aims to maintain the unlinkability of keys within its dedicated PK set, while the network needs reliable trip data for mobility pattern extraction. IoME addresses this challenge through transactions that are signed by charging stations ascertaining the presence of the EV's at their location at a specific time. These transactions from two or more stations are concatenated by verifiers to infer trips, without publicly revealing the links between the used dedicated keys of the EV. The next section illustrates this feature further.

\section{An Example Scenario}
\begin{figure*}
    \centering
    \includegraphics[width=13cm ,height=10cm,keepaspectratio]{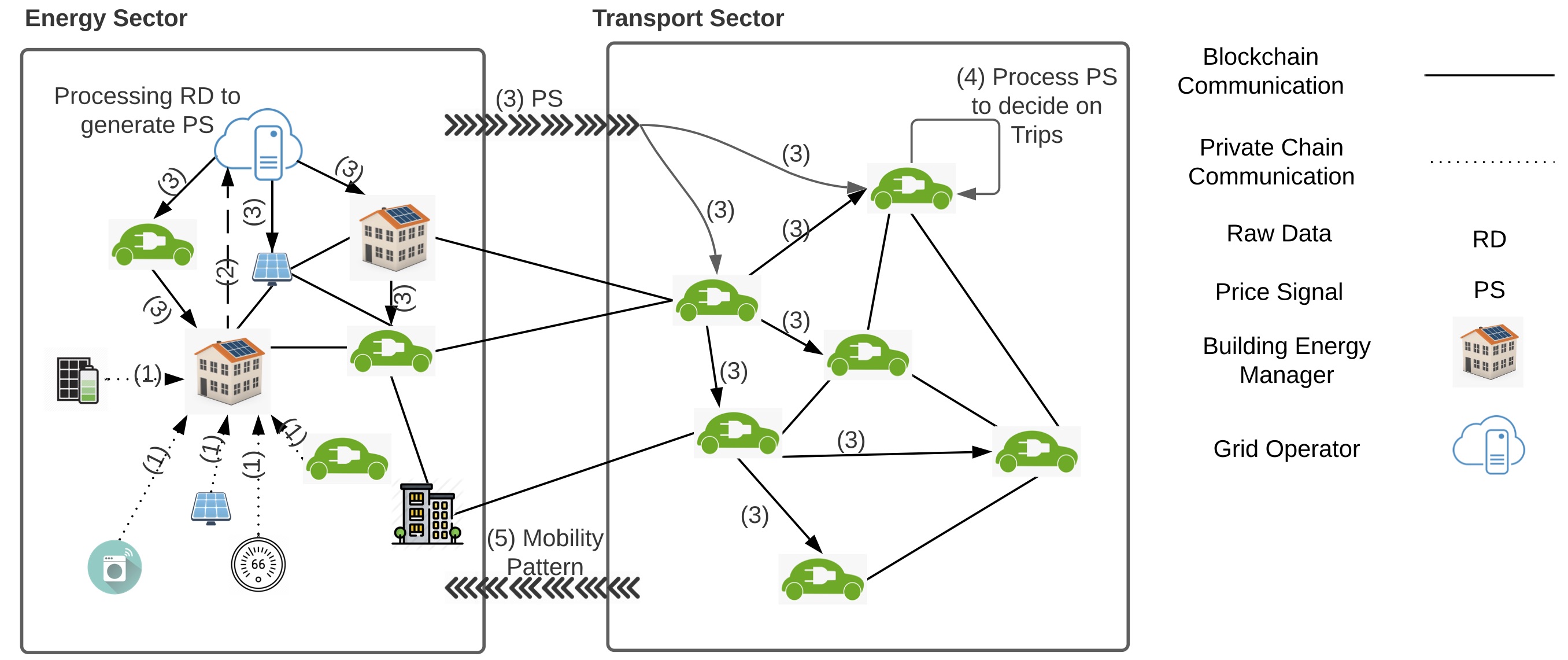}
    \caption{An overview of the proposed framework.}
    \label{fig:my_label}
\end{figure*}

In this section we outline the flow of interactions in the proposed framework through an example scenario as shown in the network topology in Fig. \ref{fig:my_label}. A public chain is managed by all the participating nodes. A private chain is run in the Virtual Power Plants (VPPs), microgrids, and the ride sharing fleet. To avoid the single point of attack, every private chain has at least one backup manager to interface with the public chain, where the backup manager can act as the manager in case the main manager is unavailable. The private chain manager is responsible for aggregating data from all private chain participants and creating transactions to be stored on the public chain. A private chain participant can store its raw data locally, and the hash of this data is stored on the private chain  for dispute resolution and cross-verification of transactions on the public chain when needed.

As outlined in Section~\ref{Sec:EnergyMobility}, the first phase in the proposed framework is to predict the future energy demand and supply. In this example scenario, the learning occurs at edge devices. 

%\begin{figure*}
  %  \centering
 %   \includegraphics[width=10cm ,height=6cm,keepaspectratio]{Mobileenergy.jpeg}
 %   \caption{An example scenario.}
 %   \label{fig:example_scenario}
%\end{figure*}

 In the energy grid, the first step (step 1 in Fig \ref{fig:my_label}) is the prediction of energy supply and demand.   We assume that each smart home is equipped with a building energy manager (BEM) which  can be integrated with the home internet gateway.  In the first learning process, the smart home energy manager collects the data related to future energy generation  and consumption which includes: solar panel, weather prediction, smart appliances, IoT devices including smart thermostat and temperature sensors, EV, and local batteries (if any). To protect the security of the communications, the data is encrypted with the PK of the BEM. Each device signs the hash of the exchanged data that protects data integrity. The participating nodes in the smart home are trusted entities and thus there is no private chain in the smart home. The BEM may receive information from third parties, e.g., weather prediction. To ensure the integrity of the data and protect against non-repudiation attack where a node denies its previous communication, the communications with external parties are stored in the public blockchain in the form of transactions.

The BEM runs the ML algorithm on the received data to predict the future energy generation and supply. The raw data of the devices is stored in a local storage along with the learning outcome. Recall that sending the raw data of the devices in the smart home increases the packet and processing overhead and also risks user privacy, thus the  BEM sends only the outcome of the learning algorithm to the public chain (step 2). To share this outcome, the BEM broadcasts an \textit{energy supply demand prediction (ESDP)} transaction structured as: 

$ T\_ID | ESDP | timestamp | PK | Sign $  

\noindent where $ T\_ID $ is the identifier of the current transaction which essentially is the hash of the transaction,  ESDP is the outcome of the learning algorithm, $timestamp$ refers to the time when the transaction is generated and the last two fields are PK and signature of the BEM.  The energy distribution companies in the public chain run the learning algorithm after receiving data from all the participants and announce the outcome using a ESDP transaction. The energy distribution company then decides on the energy price and announces that using an \textit{energy price signal (EPS)} for each region (step 3)  that is structured as follows:

$ T\_ID | P\_T\_ID | Energy\_Price | Expiry $

\noindent where $P\_T\_ID $ is the ID of the previous transaction generated by the same user. The pointer to the previous transaction ID ensures the authenticity of the transaction owner.  In public blockchains, as the users are anonymous and there is no trust, only the participants that own the key pairs corresponding to a transaction in the blockchain can create the transaction and chain it to their previous transaction. This protects against Sybil attacks where a malicious node may attempt to flood the network by sending fake transactions.  $Energy\_Price$ refers to the price of the energy and Expiry refers to the duration of time in which the price is valid for.

We next describe the information flow within the mobility grid.  When an EV receives the EPS (step 3), it first checks if there is a trip scheduled during the EPS expiry time (step 4). If so, the EV explores the possibility of opportunistically maximizing user benefits by charging the vehicle in one place and discharging in another place. The EV must consider the required energy to travel and the time to charge/discharge by comparing the net gain to move to location $L$ for discharging. The net gain is computed as the difference between the \emph{expected gain from energy discharging}, which is the difference between the feed-in tariff at location $L$ and the feed-in tariff at the vehicle's current location, and the \emph{cost of relocation} from the current location to $L$. This cost includes the direct cost to physically transport the vehicle and the depreciation cost associated with additional mileage and with reductions in the remaining recharge cycles for the EV battery. 

In case there is no travel scheduled, the EV considers the best time to charge or discharge the vehicle. Similarly, EV's used by ride sharing or other transport operators can consider energy price-driven routing decisions that consider the fleet-wide energy costs of specific transport routes, against the expected gains from energy discharging into the grid along selected routes. Both types of decisions will rely on existing mobility models~\cite{Noulas12}, the fusion of available mobility data~\cite{Liebig19}, and traffic demand prediction models~\cite{Yao18} and the SOH of this and other EV batteries, as inputs. At each time step, the EV mobility decisions will determine the charging and discharging outcomes across the transport grid, which will then be fed back into the energy demand and supply respectively for the next round of energy supply and demand prediction (step 5). 

\begin{figure*}
    \centering
    \includegraphics[width=13cm ,height=10cm,keepaspectratio]{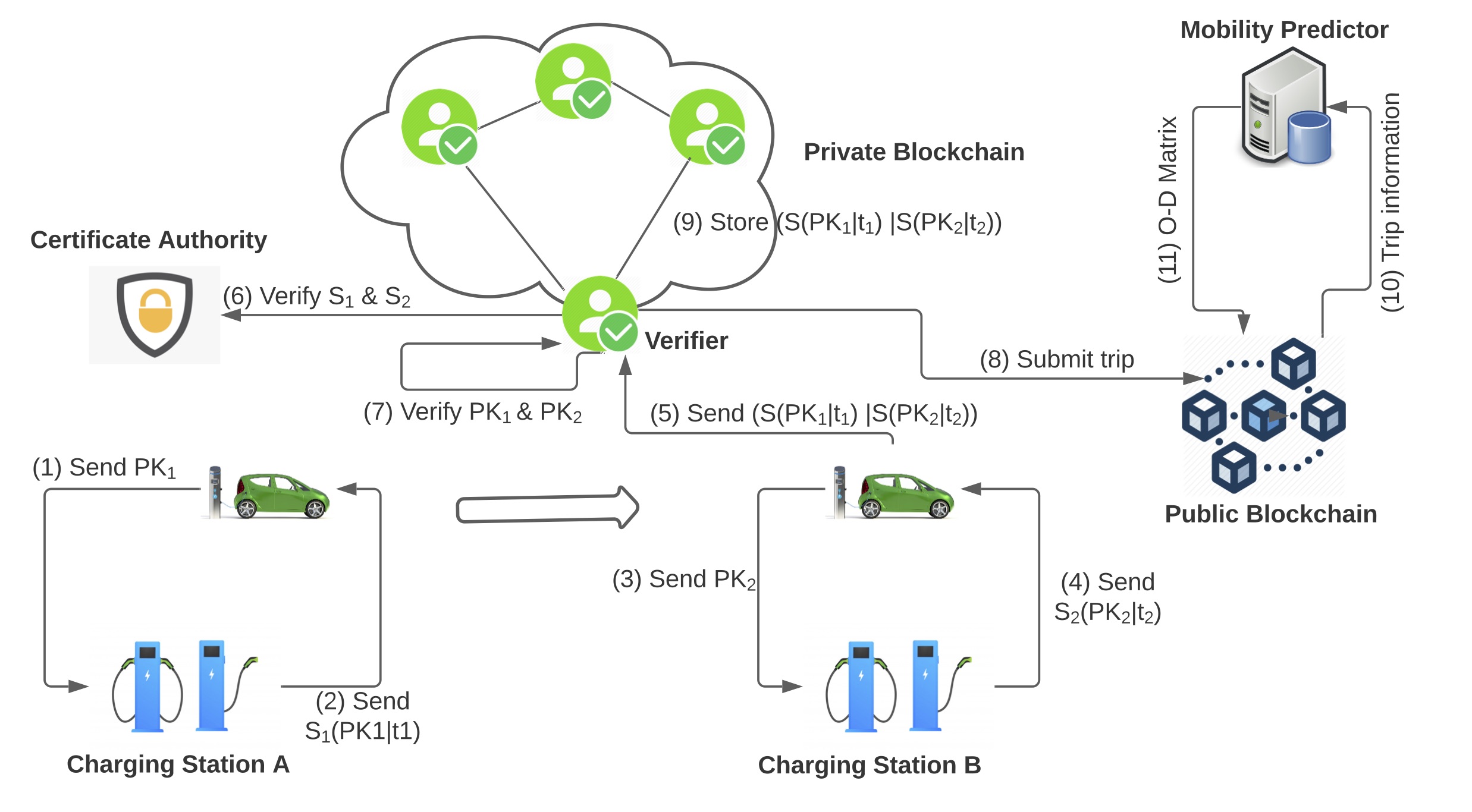}
    \caption{Anonymous trip verification.}
    \label{fig:trip-verification}
\end{figure*}

Once EV's decide on their trips, that are dilineated by source and destination charging stations, we need to anonymously extract the trip information to feed into construction of an O-D matrix. The O-D matrix feeds into the prediction of future mobility flows, which in turn is input into the next cycle of energy predictions.  Fig \ref{fig:trip-verification} depicts the process of anonymous trip verification.

To maintain its anonymity, an EV employs two one-time public keys $PK_1$ and $PK_2$ for use  at the source $A$ and destination $B$ charging stations respectively. Upon receipt of $PK_1$ (step 1 in Fig \ref{fig:trip-verification}),  $A$ signs the hash of  ($PK_1$,$t_1$), and creates  a source transaction with the following fields (step 2):\\
$ T\_ID \ |\ PK_1\ |\ Sign_A(h(PK_1,t_1) \ |\ Sign_A(PK_1)\ $\\
where $Sign_A$ indicates the signature of $A$. A similar destination transaction is created at $B$ with $PK_2$ and $t_2$ once the vehicle completes a trip (steps 3\&4). The source and destination transaction are sent to a set of verifier nodes to extract the EV's trip, without revealing the link between $PK_1$ and $PK_2$ for this EV on the public chain (step 5). The verifiers first verify  if the entity that signed the transactions is a genuine charging station by querying a Certificate Authority (step 6). The verifier then matches $PK_1$ and $PK_2$ with their corresponding signatures that ensures only the node that knows the corresponding signature can generate transactions (step 7). The extracted trip and its start and end times are anonymously stored on the public chain (step 8), which ensures the trip is recorded for generation of the O-D matrix. The individual source and destination transactions are stored in a private consortium chain that is managed by the verifiers, for future auditability of verifier actions (step 9). A mobility predictor node processes the trip information stored in the blockchain (step 10) and outputs the O-D matrix. The latter is stored in the public blockchain (step 11) and is used as input to the energy price prediction algorithm. 

\section{Discussion and Conclusion}
The IoME vision opens up several opportunities and  research questions. The energy price signal is used by the participants in the transportation industry that includes EVs and ride sharing SPs, that introduces the following benefits: i) energy transportation: each EV is equipped with a battery and thus the vehicles can be charged in one place and discharge in another place that results in purposeful energy transportation. This  is critical when the load in a particular area is increasing that may lead to the deterioration of power quality of the distribution system, ii) identifying a route toward a destination: where an EV or fleet of EVs can make routing decisions that favour cheaper energy charging costs, in addition to shorter distances and lower congestion, iii) ride sharing services: the SPs  may use the energy price signals to identify opportunities for participating drivers to earn additional income, either by choosing routes that reduce the cost of battery charging or that contribute to profitable energy transportation.

An Internet of Mobile Energy is likely to result in monetisation and marketisation of renewable energy sources, such as sunshine. Despite the increase in penetration of rooftop solar, where in Australia for instance it has reached a penetration rate of over 25\%, excess solar energy can easily reach zero or even negative price, representing a liability for energy producers. With the highly dynamic and market-driven nature of IoME, over-supply of energy in one area may be exploited by electric vehicles that opportunistically charge at times of excess supply to secure a profit by discharging later at a time of higher energy demand. Such a marketisation of renewable energy sources benefits mobility grid participants, and increases the utilisation of otherwise wasted renewable energy, and increases benefits for energy producers that would otherwise lose the value of their excess energy. 

%Transferring energy to rural areas is highly costly for energy managers particularly in remote areas due to significant cost for transmission infrastructure. Additionally, in case of natural disasters, these areas may get disconnected from the grid. The seamless transportation of energy offered by IoME enables the participants to transfer energy produced by DES and thus prevents energy shortage or disconnection in such areas.

Current EV charging and discharging technology is time-consuming, although there has been a recent trend towards fast-charging and discharging~\cite{Morrissey2016}. The vehicle must be connected to the charging station for the charging/discharging time. In recent years, charging lane roads have been developed that enable the vehicles to charge and discharge while driving which in turn impacts this delay and eliminates the need to stop in a charging station. An interesting research direction is to model this delay into routing decisions, where solutions to this problem could build on the body of knowledge in communication network routing algorithms that consider various delays in finding the optimal routes to a destination. 

Another open issue is how to conduct efficient and privacy-preserving search within IoME. For instance, a search engine may be developed for real-time location of vehicles and charging stations with attractive energy prices. The search engine needs to then include attribute-based encryption that reveals sufficient information about a node without jeopardising its privacy-sensitive information. Routing within this network can be anonymous, for instance using our anonymous backbone routing protocol~\cite{Dorri_SPB}.

%The transportation industry is also expecting disruption by the penetration of  autonomous self-driving vehicles that employ a broad range of sensors installed in the vehicle along with decision making algorithms to enable self-driving feature and eliminate the need for human interactions. The vehicles are normally parked in a location, e.g., when the owner is at work. During such time, autonomous vehicles can transport energy within IoME which in turn increases the benefits of the users. 

An important factor for IoME is to consider the energy availability from EV's and the system's sustainability, which are both determined by EV's decisions to actively participate in energy transactions. Depending on the perceived benefits and costs, EV owners will decide whether to participate.  This highlights the importance of developing an appropriate stochastic model to represent the EV participation or their availability. Such a model requires consideration of economic and behavioural factors, such as the EV owner's flexibility in adapting routing, and the incentives to drive desirable energy and transport grid outcomes. 

%Another open issue is energy provenance, which refers to ascertaining the production source of energy as green or grey, for instance. This is important as green energy currently holds a premium in smart grids, and it is not straightforward to confirm if an EV's stored energy is indeed from green sources.

An Internet of Mobile Energy will support greater convergence and seamless interaction across the transport and power sectors. IoME benefits from the maturation of technologies in both sectors and to the availability of information technologies such as IoT, blockchain, and machine learning.  While realising IoME involves several open challenges that we have outlined in this paper, an Internet of Mobile Energy is expected to deliver significant benefits for innovation, efficiency in both transport and power sectors, and welfare to end users across both sectors.

\vspace{4cm}
\begin{IEEEbiographynophoto}{Raja Jurdak}
is a Professor \& Chair at QUT. He holds a PhD degree from the UC, Irvine.  
%His research interests include trust, mobility and energy-efficiency in networks. Prof. Jurdak has published over 180 peer-reviewed publications, including two authored books. He serves on the editorial board of Ad Hoc Networks and on the organising and technical program committees of top international conferences, including Percom, ICBC, and IPSN. He is a conjoint professor with the University of New South Wales, a visiting scientist at CSIRO, and a senior member of the IEEE.
\end{IEEEbiographynophoto}
\vspace{-6cm}
\begin{IEEEbiographynophoto}{Ali Dorri}
is a Research Fellow at QUT. He received his Ph.D. degree from UNSW. 
%He was also a Postgraduate research student at CSIRO. His research interest includes blockchain, Internet of Things (IoT), security and privacy, and distributed systems. He has published over 25 peer-reviewed papers. Ali served on the organizing committee of SDLT and BCCA and as technical program committee in 10 conferences including ICBC.
\end{IEEEbiographynophoto}
\vspace{-6cm}
\begin{IEEEbiographynophoto}{D. Mahinda Vilathgamuwa}
received %his B.Sc. degree in electrical engineering from the University of Moratuwa, Sri Lanka, in 1985, and 
the Ph.D. degree from Cambridge University. 
%He joined  Nanyang Technological University in 1993 and served as a faculty member until 2013. He 
He is currently a Professor at QUT. 
%He has published over 300 research papers in refereed journals and conferences, 1 book and 2 book chapters. He was an associate editor for IEEE Transactions on Industry Applications from 2014 to 2017 and 
%He currently serves as an associate editor for IEEE Transactions on Industrial Electronics. He is an IEEE fellow.
\end{IEEEbiographynophoto}
%\enlargethispage{-9.5cm}
\end{document}